\documentclass[12pt]{article}

\usepackage[tbtags]{amsmath}
\usepackage{amsmath}
\usepackage{amssymb}
\usepackage{amsthm}

\newcommand{\pde}{partial differential equation}
\newcommand{\ode}{ordinary differential equation}
\newcommand{\gen}[1]{\partial_{#1}}
\newcommand{\dil}{x\gen x+y\gen y+t\gen t}

\newcommand{\curl}[1]{ \{#1\} }

\DeclareMathOperator{\SL}{SL}
\DeclareMathOperator{\sn}{sn}
\DeclareMathOperator{\cn}{cn}
\DeclareMathOperator{\dn}{dn}

\newcommand{\R}{\mathbb{R}}

\numberwithin{equation}{section}
\begin{document}
\title{Exact Solutions of a (2+1)-Dimensional\\ Nonlinear
Klein-Gordon Equation}
\author{F. \textsc{G{\"u}ng{\"o}r}\\
\small
\begin{tabular}{c}
Department of Mathematics, Faculty of Science\\
Istanbul Technical University\\
80626, Istanbul, Turkey\\
\texttt{gungorf@itu.edu.tr}
\end{tabular}}
\date{}
\maketitle
\begin{center}
\end{center}

\abstract{The purpose of this paper is to present a class of particular
solutions of a C(2,1) conformally invariant nonlinear Klein-Gordon
equation by
symmetry reduction. Using the subgroups of similitude group reduced
ordinary differential equations of second order and their solutions by a
singularity analysis are classified. In particular, it has been shown
that  whenever they have the
Painlev\'e property,  they can be transformed to standard forms by
Moebius transformations of dependent variable and arbitrary smooth
transformations of independent variable whose solutions, depending on the
values of parameters, are expressible in terms
of  either elementary functions or Jacobi elliptic functions.}
\section{Introduction}
In  a recent paper \cite{Gungor98}, we constructed second order differential equations
invariant under the Poincar\'e, similitude and conformal groups in
$(2+1)$-dimensional space-time. For instance, the planar nonlinear
Klein-Gordon (NLKG) equation
$$\square_3 u=H(u),$$
where  $H$ is an arbitrary sufficiently smooth function of its argument and
$\square_3=\partial_{t}^{2}-\Delta_2$ is the wave operator in the
(2+1)-dimensional Minkowski space,
is invariant under Poincar\'e group P(2,1) and more specifically, NLKG
equation with a power nonlinearity
\begin{equation}\label{1.1}
\square_3 u=a u^k
\end{equation}
is invariant under the Poincar\'e group extended by dilations, also called
similitude group. Among equations of the form \eqref{1.1}, the special case $k=5$ plays a
privileged role in classical  and quantum field theory.
In this context Eq. \eqref{1.1} with $k=5$, namely
\begin{equation}\label{1.2}
\square_3 u=a u^5
\end{equation}
where $a$ is an arbitrary constant,
arises as the equation of motion (Euler-Lagrange equation)
obtained minimizing the action corresponding to a Lagrangian density
$${\cal L}=1/2(\nabla u)^2-6 a u^5=1/2(u_t^2-u_x^2-u_y^2)-6a u^5.$$
This equation is also called classical $\phi^6$-field equation.
For further physical motivation of this equation, the reader is referred to
Ref. \cite{winternitz1}
Another remarkable property  of   Eq. \eqref{1.2} is that,
in addition to being similitude invariant, it is also
invariant under the conformal group C(2,1) of space-time.

It will be of interest to find  exact solutions of the NKLG equation
\eqref{1.2},
not only from  mathematical but also from physical point of view. 
While there exists an extensive literature on exact solutions of the
NLKG equation they are mainly devoted to translation invariant solutions
in 1+1 dimension. The study of  exact solutions in
higher dimensions  appears in a few papers only. For example, in
\cite{winternitz1} exact solutions are studied in (3+1)-dimensional case.
Another article dedicated to A.O.Barut \cite{grundland} investigated
translationally
invariant solutions which actually live in a (1+1)-dimensional case and
static spherically symmetric and similarity solutions in 3+1 dimensions.
To our knowledge, there exists no systematic study of exact solutions of
the NKLG equation in 2+1 dimensions and the present paper aims at
obtaining  exact solutions of the  NLKG equation.

The method to be used for solving \eqref{1.2} is the symmetry method.
This method is described in various books \cite{olver1, bluman} and 
lectures \cite{winternitz2, winternitz3}.  Applications of the method to 
find exact solutions of numerous  equations of nonlinear mathematical
physics are given in \cite{Fushchich93}.
The first step is to reduce the
considered PDE to an ODE expressed in terms of symmetry variables. Next,
we integrate, whenever possible, the obtained ODEs.
For the NLKG equation \eqref{1.2}, except for
degenerate cases simplifying either to an algebraic or to a first order
equation that can be integrated directly, most of the reduced ODEs will
be second order and nonlinear.

Often, two approaches are adopted for solving these ODEs. The first 
is to find the symmetry group of the reduced equation, if one exists, and 
then using it  to lower the order of the equation.
The second one consists of  performing
a singularity analysis in order to establish whether the equation is of
the Painlev\'e type meaning that the general solution of the
corresponding equation has no movable singularities (branch points,
essential singularities)  other than the poles.
Note that the second method gives  more satisfactory results than the first
one. while the equation has no nontrivial symmetries, it may well belong to
the class of the Painlev\'e type equations. Equations of the form
\begin{equation*}\label{1.3}
w''(z)=f(z,w,w')
\end{equation*}
where $f$ is analytic in $z$,  rational in $w'$, and algebraic in $w$,
possessing the
Painlev\'e property was classified by Painlev\'e and Gambier. They
showed that such an ODE can be reduced to one of the 50 equations
listed in \cite{ince, davis} whose solutions can be expressed in terms of
either elementary or Jacobi elliptic functions or of solutions of linear
equations. Six of them are irreducible and  known to be Painlev\'e
transcendents which often occur in a host of physical problems.
Equations that do not possess the Painlev\'e property will in general
contain moving (logarithmic) singularities and we have no systematic
method for integrating them. 

In order to obtain in a systematic manner symmetry reductions of Eq.
\eqref{1.2} we need a classification of subgroups of the symmetry group
into conjugacy classes under the action of the symmetry group.
The subgroups of similitude groups are known only for small dimensions,
while those of Poincar\'e groups having generic orbits of codimension one are
known for arbitrary dimension.
The subgroups  of S(2,1) are classified in  \cite{patera6}. Making
use of this result we obtain all possible reduced ODEs, mostly of second order
nonlinear. In order to facilitate the tedious computations the MATHEMATICA
package has been used.
\section{The Symmetry Group and its  Lie Algebra}
The symmetry group of \eqref{1.2} called similitude group or 
extended Poincar\'e group S(2,1) is the group of
transformations leaving the Lorentz metric form invariant. Its structure
is
\begin{equation}\label{2.1}
S(2,1)=D\vartriangleright(\SL(2,\R)\vartriangleright T_3)
\end{equation}
where $T_3$ are space-time translations, $\SL(2,\R)$ is the special
linear group, $D$ dilations and $\vartriangleright$ denotes a
semi-direct product.
A convenient basis for the Lie algebra of the extended Poincar\'e group
is given by  translations $\curl{P_0,P_1,P_2}$, Lorentz boosts
$\curl{K_1,K_2}$, rotation $L_3$, and dilation $D$. For Eq. \eqref{1.2}
the basis is realized as
\begin{subequations}
\begin{equation}\label{p21}
\begin{array}{llll}
& P_0=\gen t,\quad P_1=\gen x, & &  P_2=\gen y,   \\
& K_1=t\gen x+x\gen t,   &  & K_2=t\gen y+y\gen t    \\
& L_3=y\gen x-x\gen y,   &  & D=\dil-u/2\gen u.
\end{array}
\end{equation}
In addition to generators \eqref{p21}, Eq. \eqref{1.2} has also
conformal symmetries generated by
\begin{equation}\label{p22}
\begin{array}{lll}
& C_0=2xt\gen x+2yt\gen y+(x^2+y^2+t^2)\gen t-tu\gen u\\
& C_1=(t^2+x^2-y^2)\gen x+2xy\gen y+2xt\gen t-ux\gen u\\
& C_2=2xy\gen x+(t^2-x^2+y^2)\gen y+2yt\gen t-uy\gen u.
\end{array}
\end{equation}
\end{subequations}

\section{Symmetry Reductions of  NLKG Equation}
In this section we give a classification of symmetry reductions of 
\eqref{1.2} with respect to  invariance under the similitude group.
Applying the method of symmetry reduction   we will derive all the
solutions invariant under subgroups with generic orbits of codimension 1
in the space of independent variables.
However for the sake of completeness, first, we obtain all reductions of
\eqref{1.2} to lower dimensional equations with  two independent
variables.
\subsection{Symmetry Reductions to  
PDEs in Two Variables}\label{pdered}
We restrict ourselves to subgroups with generic orbits of codimension
two in the space-time $(x,y,t)$ and of codimension three in the space $(x,y,t,u)$.
The corresponding three invariants $I_i(x,y,t,u),\; i=1,2,3$ of the group action
on $X\otimes U$ must provide an invertible transformation from the space of
dependent variables to that of the invariants. Hence the invariants of
the subgroup
$H$ of the symmetry group can be written in the form
\begin{equation*}
\begin{split}
I_1&=\xi (x,y,t),\qquad I_2=\eta(x,y,t),\\
I_3&=f(x,y,t,u)=u \tilde\phi(x,y,t).
\end{split}
\end{equation*}
This permits us to reduce \eqref{1.2} to a  \pde{} for $f(\xi,\eta)$ which is
a function of the symmetry variables $\xi$ and $\eta$ and write the solution of
\eqref{1.2} as
\begin{equation}\label{3.1a}
u(x,y,t)=f(\xi,\eta)\phi(x,y,t)
\end{equation}
where $\phi$, $\xi$ and $\eta$ are known functions whose precise forms are to
be determined by the choice of subgroup. Substituting the reduction
formula \eqref{3.1a} into \eqref{1.2} we obtain a \pde{} for the function $f$.

The classification of all subgroups of the similitude group is well known
\cite{patera6}.
Using these classification results we classify symmetry reductions.
In  the following we give reduction formulas and reduced equations for all
possible subgroups:
\begin{itemize}
\item[(1)] \quad Subgroup $\curl{K_2+L_3}:$
\begin{equation}\label{}
\begin{split}
&u=f(\xi,\eta), \qquad \xi=x+t, \qquad \eta=y^2-2t(x+t),\\[.5cm]
&4(\xi^2-\eta)f_{\eta\eta}-4\xi f_{\xi\eta}-6f_{\eta}+a f^5=0.
\end{split}
\end{equation}
\item[(2)] \quad Subgroup $\curl{K_1}:$
\begin{equation}\label{}
\begin{split}
&u=f(\xi,\eta), \qquad \xi=t^2-x^2,\\[.5cm]
&4\xi^2 f_{\xi\xi}-f_{yy}+4f_\xi-af^5=0.
\end{split}
\end{equation}
\item[(3)] \quad Subgroup $\curl{D}:$
\begin{equation}\label{}
\begin{split}
&u=t^{-1/2}f(\xi,\eta),\qquad \xi=\frac{y}{t},\qquad \eta=\frac{x}{t},\\[.5cm]
&4(\xi^2-1)f_{\xi\xi}+4(\eta^2-1)f_{\eta\eta}+8\xi\eta
f_{\xi\eta}\\[.5cm]
&+12\xi f_\xi+12\eta f_{\eta}+3f-4a f^5=0.
\end{split}
\end{equation}
\item[(4)] \quad Subgroup $\curl{P_2}:$
\begin{equation}\label{}
\begin{split}
&u=f(x,t),\\[.5cm]
&f_{tt}-f_{xx}-af^5=0, \qquad y\text{-independent equation}  .
\end{split}
\end{equation}
\item[(5)] \quad Subgroup $\curl{L_3}:$
\begin{equation}\label{}
\begin{split}
&u=f(\xi,t),\qquad \xi=(x^2+y^2)^{1/2},\\[.5cm]
&f_{tt}-f_{\xi\xi}-1/\xi f_{\xi}-a f^5=0.
\end{split}
\end{equation}
\item[(6)] \quad Subgroup $\curl{P_0}:$
\begin{equation}\label{}
\begin{split}
&u=f(y,t),\\[.5cm]
&f_{tt}-f_{yy}-af^5=0,\qquad x\text{-independent equation}.
\end{split}
\end{equation}
\item[(7)] \quad Subgroup $\curl{P_0-P_1}:$
\begin{equation}\label{}
\begin{split}
&u=f(\xi,y),\qquad \xi=x+t ,\\[.5cm]
&f_{yy}+bf^5=0.
\end{split}
\end{equation}
\item[(8)] \quad Subgroup $\curl{K_1+P_2}:$
\begin{equation}\label{}
\begin{split}
&u=f(\xi,\eta),\qquad \xi=t^2-x^2,\qquad \eta=(x+t)e^{-y},\\[.5cm]
&\eta^2f_{\eta\eta}-4\eta f_{\xi\eta}-4\xi^2 f_{\xi\xi}+\eta f_{\eta}
-4\xi f_{\xi}+a f^5=0.
\end{split}
\end{equation}
\item[(9)] \quad Subgroup $\curl{L_3+P_0}:$
\begin{equation}\label{}
\begin{split}
&u=f(\xi,\eta),\qquad \xi=x^2+y^2,\qquad \eta=\arctan \frac{y}{x}-t,\\[.5cm]
&4\xi^2f_{\xi\xi}+(1-\frac{1}{\xi^2})f_{\eta\eta}-4f_{\xi}-af
^5=0.
\end{split}
\end{equation}
\item[(10)] \quad Subgroup $\curl{D+b K_1}\quad 0< b\le 1:$
\begin{equation}\label{}
\begin{split}
&u=y^{-1/2}f(\xi,\eta),\qquad \xi=(x+t)y^{b-1},\qquad
\eta=(x+t)y^{-(b+1)},\\[.5cm]
& (b-1)^2\xi^2 f_{\xi\xi}-2[(1-b^2)\xi\eta-2]f_{\xi\eta}
+(b+1)^2\eta^2 f_{\eta\eta}\\[.5cm]
&+(b-1)(b-3)\xi f_{\xi}+(b+1)(b+3)\eta f_{\eta}
+3/4 f+a f^5=0.
\end{split}
\end{equation}
\item[(11)] \quad Subgroup $\curl{D+ b L_3},\quad b>0:$
\begin{equation}\label{}
\begin{split}
&u=t^{-1/2}f(\xi,\eta),\qquad \xi=(x^2+y^2)^{-b/2}e^{\arctan
y/x}, \qquad \eta=(x^2+y^2)t^{-2},\\[.5cm]
&(1+b^2)\xi^2 f_{\xi\xi}-4b \xi\eta f_{\xi\eta}-4\eta^2(\eta-1)f_{\eta\eta}\\[.5cm]
&+(1+b^2)\xi^2 f_{\xi}-4\eta (2\eta-1)f_{\eta}-(3/4)\eta f+a \eta f^5=0.
\end{split}
\end{equation}
\item[(12)] \quad Subgroup $\curl{D+K_1+P_0+P_1}:$
\begin{equation}
\begin{split}
&u=y^{-1/2}f(\xi,\eta),\qquad \xi=x-t,\qquad \eta=(1+x+t)y^{-2},\\[.5cm]
&16\eta^2  f_{\eta\eta}+16f_{\xi\eta}+32\eta f_{\eta}+3f+4 a f^5=0.
\end{split}
\end{equation}
\end{itemize}
\subsection{Symmetry Reductions to Ordinary Differential
Equations}\label{odered}
Subgroups of the symmetry group that have generic orbits of codimension one in
the space of independent variables $(x,y,t)$ and of codimension two in
$(x,y,t,u)$ space will provide reductions to \ode{}. The invariants of subgroup
$H$  have the form
$$I_1=\xi(x,y,t)\qquad \text{and} \qquad
I_2=f(x,y,t,u)=\tilde\phi(x,y,t) u.$$
In this case the reduction formula will be
$$u(x,y,t)=\phi(x,y,t)f(\xi)$$ where $\phi$ and $\xi$ are again known functions.
If the action of the subgroup $H$, restricted to the time-space variables, is
transitive the \ode{} is of first order, otherwise of second order. 
We mention that subgroups with codimension zero in the time-space variables
will reduce the original equation to an algebraic equation that may or may not
admit nontrivial solutions. We run
through the individual subgroups and obtain the following reductions:
\begin{itemize}
\item[(1)] \quad Subgroup $\curl{P_1,D+b K_2}:$
\begin{equation}\label{ode1}
\begin{split}
&u=(t^2-y^2)^{-1/4}f(\xi),\qquad \xi=(t-y)^{b+1}(t+y)^{b-1},\\[.5cm]
&16(b^2-1)\xi^2f''+8(2b^2-b-2)\xi f'+f-4a f^5=0.
\end{split}
\end{equation}
\item[(2)] \quad Subgroup $\curl{D+b L_3,P_0}:$
\begin{equation}\label{ode2}
\begin{split}
&u=(x^2+y^2)^{-1/4}f(\xi),\qquad
\xi=\arctan{y/x}-b/2\log(x^2+y^2),\\[.5cm]
&4(1+b^2)f''+4bf'+4af^5+f=0.
\end{split}
\end{equation}
\item[(3)]\quad Subgroup $\curl{D+b K_1,P_0-P_1},\quad -1<b\le 1,\quad b\ne 0:$
\begin{equation}\label{ode3}
\begin{split}
&u=y^{-1/2}f(\xi),\qquad \xi=(x+t)y^{-(b+1)},\\[.5cm]
&b^2\xi^2f''+b(b+2)\xi f'+3/4 f+a f^5=0.
\end{split}
\end{equation}
\item[(4)] \quad Subgroup $\curl{D+K_1+P_0+P_1,P_0-P_1}:$
\begin{equation}\label{ode4}
\begin{split}
&u=y^{-1/2}f(\xi),\qquad \xi=(x+t+1)y^{-2},\\[.5cm]
&16\xi^2 f''+32 \xi f'+3f+4a f^5=0.
\end{split}
\end{equation}
\item[(5)] \quad Subgroup $\curl{D+K_2+P_0+P_1,P_1}:$
\begin{equation}\label{ode5}
\begin{split}
&u=(t+y+1)^{-1/4}f(\xi),\qquad \xi=t-y,\\[.5cm]
&f'+a f^5=0.
\end{split}
\end{equation}
\item[(6)] \quad Subgroup $\curl{D+b K_1,K_2+L_3},\quad b>0:$
\begin{equation}\label{ode6}
\begin{split}
&u=(t^2-x^2-y^2)^{-1/4}f(\xi),\qquad \xi=(t^2-x^2-y^2)^{b+1}(t+x)^{-2},\\[.5cm]
&4(b^2-1)\xi^2f''+2(2b^2-1)\xi f'-1/4 f-a f^5=0.
\end{split}
\end{equation}
\item[(7)] \quad Subgroup $\curl{D-K_1+P_0-P_1,K_2+L_3}$
\begin{equation}\label{ode7}
\begin{split}
&u=(x^2+y^2-t^2-x-t)^{-1/4}f(\xi),\qquad \xi=x+t,\\[.5cm]
&4\xi f'+f-4a f^5=0.
\end{split}
\end{equation}
\item[(8)] \quad Subgroup $\curl{D,P_0}:$
\begin{equation}\label{ode8}
\begin{split}
&u=t^{-1/2}f(\xi),\qquad \xi=y/t,\\[.5cm]
&4(1-\xi^2)f''-12 \xi f'-3 f+4a f^5=0.
\end{split}
\end{equation}
\item[(9)] \quad Subgroup $\curl{P_0-P_1,D}:$
\begin{equation}\label{ode9}
\begin{split}
&u=y^{-1/2}f(\xi),\qquad \xi=(x+t)/y,\\[.5cm]
&4\xi^2 f''+12\xi f'+3f+4a f^5=0.
\end{split}
\end{equation}
\item[(10)] \quad Subgroup $\curl{D,P_2}:$
\begin{equation}\label{ode10}
\begin{split}
&u=t^{-1/2}f(\xi),\qquad \xi=x/t,\\[.5cm]
&4(1-\xi^2)f''-12\xi f'-3 f+4a f^5=0.
\end{split}
\end{equation}
\item[(11)] \quad Subgroup $\curl{K_1,K_2+L_3}:$
\begin{equation}\label{ode11}
\begin{split}
&u=f(\xi),\qquad \xi=x^2+y^2-t^2,\\[.5cm]
&4\xi f''+6f'+a f^5=0.
\end{split}
\end{equation}
\item[(12)] \quad Subgroup $\curl{P_0-P_1,K_1+P_2}$
\begin{equation}\label{ode12}
\begin{split}
&u=f(\xi),\qquad \xi=e^{-y}(x+t),\\[.5cm]
&\xi^2 f''+\xi f'+a f^5=0.
\end{split}
\end{equation}
\item[(13)] \quad Subgroup $\curl{P_0-P_1,D+K_2+L_3}:$
\begin{equation}\label{ode13}
\begin{split}
&u=(x+t)^{-1/2}f(\xi),\qquad \xi=\frac{y}{x+t}-\log (x+t),\\[.5cm]
&f''+a f^5=0.
\end{split}
\end{equation}
\item[(14)] \quad Subgroup $\curl{K_1,K_2,L_3}:$
\begin{equation}\label{ode14}
\begin{split}
&u=f(\xi),\qquad \xi=t^2-x^2-y^2,\\[.5cm]
&4\xi f''+6 f'-a f^5=0.
\end{split}
\end{equation}
\item[(15)] \quad Subgroup $\curl{L_3,P_1,P_2}:$
\begin{equation}\label{ode15}
u=f(t),\qquad f''-a f^5=0.
\end{equation}
\item[(16)] \quad Subgroup $\curl{K_1,P_0,P_1}:$
\begin{equation}\label{ode16}
u=f(y),\qquad f''+af^5=0.
\end{equation}
\end{itemize}
\section{Discussion of the Reduced Ordinary Differential
Equations}\label{trans}
Once the reduced ODEs were obtained the remaining task will be to transform
them,
whenever they have the Painlev\'e property, into one of the  standard forms  that
can be integrated once with the exception of the Painlev\'e transcendents.
By a rescaling of  independent and dependent variables
all second order ODEs obtained through the symmetry reduction can be
written in a unified manner as
\begin{equation}\label{odemain}
A(\xi) f''(\xi)+B(\xi) f'(\xi)+C(\xi)f(\xi)+D(\xi) f^5(\xi)=0.
\end{equation}
We now pick out those having the Painlev\'e property. To achieve this
task we subject the reduced equations to the Painlev\'e test which provides
the necessary conditions for having the Painlev\'e property. 
Eq. \eqref{odemain} itself does not directly have the Painlev\'e property.
However, a leading order analysis indicates that if we  make the 
substitution $$f(\xi)=\sqrt{h(\xi)}, \quad h(\xi)> 0$$
then the equation for $h$ may have the Painlev\'e property.

In the following we run through all the ODEs separately :
\subsection*{Equation \eqref{ode1}:}
\begin{equation}\label{eq1}
B\xi^2 f''+A \xi f'+f^5=0\qquad B=16(b^2-1),\; A=8(2b^2-b-2).
\end{equation}
By a change of independent variable $\eta=\ln \xi$ it reduces to
\begin{equation}\label{red}
B \ddot f+(A-B)\dot f+f-4a f^5=0
\end{equation}
where dot denotes derivative with respect to $\eta.$
For $b=\pm 1$ after scaling variables we have
\begin{equation}\label{eq1a}
\dot f=f(f^4-1).
\end{equation}
Its solution with the original variable is
$$f=(1-\xi_0 \xi)^{-1/4}$$
where $\xi_0$ is an integration constant.
For $b=0$, it has the form
\begin{equation}\label{eq1b}
f''+f+f^5=0.
\end{equation}
This equation passes the Painlev\'e test.
For $b\ne \pm 1$,
Putting
$$z=f\qquad w(z)=\dot f$$ transforms equation \eqref{red} to
$$16(b^2-1)w w_z-8b w+z-4a z^5=0$$
which is an Abel equation of the second kind. Some remarks on this equation
are noteworthy. This equation is not
tractable by standard methods, meaning that there is no systematic method
for solving it in closed form, neither a substitution transforming it into
a linear equation.
A list of  solvable examples  can be found in the collection of
\cite{kamke}. In a very recent paper
\cite{Schwarz98}, F. Schwarz studied symmetry analysis of Abel equation and
showed that, when the coefficients of the rational normal form of the 
equation satisfy some constraint,
Abel equation admits a one-parameter structure-preserving symmetry group
reducing it to a quadrature. Existence of a two-parameter symmetry group
implies that equation is actually equivalent to a Bernoulli equation.

\subsection*{Equation \eqref{ode2}:}
\begin{equation}\label{eq2}
f''+A f'+f+f^5=0, \quad A=2b(1+b^2)^{-1/2}.
\end{equation}
Eq. \eqref{eq2} passes the Painlev\'e test only for $b=0 (A=0)$. For $b\ne 0$,
a transformation from $(\xi,f(\xi))\to (z,w(z))$ by setting $z=f,\quad w(z)=f'$
brings \eqref{eq2} to an Abel equation of the second kind
$$w w_z+A w+z+z^5=0.$$
\subsection*{Equations \eqref{ode3}, \eqref{ode4}, \eqref{ode6}, \eqref{ode9}:}
They are all treated as similar to  equation \eqref{ode1}.
\subsection*{Equation \eqref{ode5}:}
Eq. \eqref{ode5} is immediately integrated to give the singular solution
$$f=\curl{4a(\xi_0-\xi)}^{-1/4}$$
where $\xi_0$ is an integration constant.
\subsection*{Equation \eqref{ode7}:}
\begin{equation}\label{eq7}
4\xi f'+f-4a f^5=0.
\end{equation}
Let us transform the independent variable from $\xi$ to $\zeta=\ln \xi$.
Scaling  variables lead to
$$f_\zeta=f(f^4-1)$$
which is  equation \eqref{eq1a} again.
\subsection*{Equation \eqref{ode8}:}
\begin{equation}\label{eq8}
(1-\xi^2)f''-3\xi f'-\frac{3}{4}f+f^5=0.
\end{equation}
This equation has the Painlev\'e property. Eq. \eqref{ode10} is treated
similarly.
\subsection*{Equation \eqref{ode11}:}
\begin{equation}\label{eq11}
\xi f''+\frac{3}{2}f'+f^5=0.
\end{equation}
This equation has the Painlev\'e property.
Eq. \eqref{ode14} is treated similarly .
\subsection*{Equations \eqref{ode12}, \eqref{ode13}, \eqref{ode15},
\eqref{ode16}:}
All of these equations can be transformed to
\begin{equation}\label{eq235}
f''+f^5=0
\end{equation}
which has  the first integral
$${f'}^2+1/3 f^6=C.$$
where $C$ is an  integration constant.
Setting $\phi=f^2$ this equation is further transformed to
the elliptic function equation
$${\phi'}^2=4C\phi-4/3\phi^4$$
The form of solutions will depend on the values of $C$.  The solutions of
a more general equation of this type will be discussed below.

In summary, among the considered equations the only ones that pass the
Painlev\'e test are
\begin{subequations}
\begin{equation}\label{p1}
f''+f+f^5=0,
\end{equation}
\begin{equation}\label{p2}
(1-\xi^2)f''-3\xi f'-\frac{3}{4}f+f^5=0,
\end{equation}
\begin{equation}\label{p3}
\xi f''+\frac{3}{2}f'+f^5=0.
\end{equation}
\end{subequations}
The substitution $f=\sqrt{h}$ transforms these equations to
\begin{subequations}
\begin{equation}\label{pp1}
h''=\frac{{h'}^2}{2h}-2(h^3+h),
\end{equation}
\begin{equation}\label{pp2}
h''=\frac{{h'}^2}{2h}-\frac{1}{2(1-\xi^2)}(4h^3-3h-6\xi h'),
\end{equation}
\begin{equation}\label{pp3}
h''=\frac{{h'}^2}{2h}-\frac{1}{\xi}(2h^3+\frac{3}{2}),
\end{equation}
\end{subequations}respectively.
By a rescaling of variables \eqref{pp1} can be brought to a
standard form \cite{ince}
\begin{equation}\label{standard}
h''=\frac{{h'}^2}{2h}+\frac{3}{2}h^3-\frac{h}{2}
\end{equation}
which has the first integral
\begin{equation}\label{first}
{h'}^2=h^4-h^2+C h=h(h^3-h+C),
\end{equation}
where $C$ is an integration constant. Solutions of this equation can be
expressed in terms of Jacobi elliptic functions depending on the value of $C$.
By a linear transformation
$$h(\xi)= \frac{3}{2}(1-\xi^2)^{-1/2} W(\eta),\quad
\eta=\ln(\xi+\sqrt{\xi^2+1})$$
the standard form of \eqref{pp2} becomes \eqref{standard}. Finally,
transforming \eqref{pp3} by
$$h(\xi)=\sqrt{-\frac{16}{3}}\xi^{-1}W(\eta), \quad
\eta=-\frac{16}{3}\xi^{-3/2}$$ we have again \eqref{standard}.
We rewrite Eq. \eqref{first}  as
\begin{equation}\label{elliptic}
{h'}^2=P(h)=h(h-h_1)(h-h_2)(h-h_3)
\end{equation}
with
\begin{equation*}
\begin{array}{ll}
& h_1+h_2+h_3=0  \\[2mm]
& h_1 h_2+h_1 h_3+h_2 h_3=-1  \\[2mm]
& h_1 h_2 h_3=-C.
\end{array}
\end{equation*}
The above equation can be simplified by a Moebius transformation of the
dependent variable
$$h(\xi)=\frac{\rho Z(\xi)+\sigma}{\mu Z(\xi)+\nu},$$
where $\rho, \sigma, \mu, \nu$ are constants. If all four roots of
$P(h)$ are distinct we choose $\rho, \sigma, \mu, \nu$ so as to to
transform the zeros at $0$, $h_1$, $h_2$ and $h_3$ into zeros at $\pm 1$
and $\pm M$, where $M$ is some constant. In other words, we transform
\eqref{elliptic} into the standard form
\begin{equation}\label{standard-elliptic}
{Z'}^2=K (1-Z^2)(M^2-Z^2).
\end{equation}
The form of solution of this
equation
depends on the multiplicity of the roots of the polynomial $P(h)$.
Solutions can be real or complex, finite or singular, periodic or
localized.
If the four roots are distinct, we obtain solutions in terms of Jacobi
elliptic functions.
The general solution of \eqref{standard-elliptic} is given by
\begin{equation*}
\begin{array}{ll}
Z&=\sn (\sqrt{K}M(\xi-\xi_0),M^{-1}),\quad \text{for}\quad M^2>1,\; M^2\in
\R\\[2mm]
Z&=\cn (\sqrt{-K(1-M^2)}(\xi-\xi_0),(1-M^2)^{-1/2}),\quad \text{for}\quad
M^2<0\\[2mm]
Z&=\dn (\sqrt{-K}(\xi-\xi_0),(1-M^2)^{1/2}),\quad \text{for}\quad 0<M^2<1.
\\[2mm]
\end{array}
\end{equation*}
They are hence always periodic rather than
localized, and they may be finite or have periodically spaced
singularities (poles) on the real axis.
A detailed treatment  of these functions can be found in any book on
elliptic functions (for example, see Ref. \cite{byrd}).

If any of the roots have
multiplicity higher than one, we obtain elementary solutions. These can
be localized, namely solitary waves or kinks. They can also be periodic
and hence delocalized. They can have singularities on the real axis.
Some elementary solutions of \eqref{elliptic} are listed below:
\begin{equation*}
\begin{array}{ll}
a.)& h_1=h_2=h_3=0\\[2mm]
   & h=(\xi-\xi_0)^{-1}\\[2mm]
b.)& h_2=h_3=0,\; h_1\ne 0\\[2mm]
   & h=h_1\curl{1-h_1^2(\xi-\xi_0)^2/4}^{-1}\\[2mm]
c.)& h_1=h_2=h_3\ne 0\\[2mm]
   & h=h_1(\xi-\xi_0)^2\curl{(\xi-\xi_0)^2-4h_1^{-2}}^{-1}\\[2mm]
d.)& h_3=0,\; h_1=h_2\ne 0\\[2mm]
   & h=h_1[1-e^{(\xi-\xi_0)}]^{-1}\\[2mm]
e.)& h_3=0,\; h_1\ne h_2\ne 0\\[2mm]
   & h=h_1 h_2\curl{(h-h_1)\cosh \sqrt{h_1h_2}(\xi-\xi_0)+h_1+h_2}^{-1}
   \\[2mm]
f.)& h_2=h_3\ne 0, \; h_1\ne h_3\ne 0\\[2mm]
   & h=h_1h_2 \tanh^2 X\curl{h_1-h_2+h_2\tanh^2 X}^{-1},\quad
X=\sqrt{h_2(h_2-1)}/2(\xi-\xi_0).
\end{array}
\end{equation*}

In addition to similitude invariant exact solutions, conformal
transformations generated by \eqref{p22} can be used to obtain conformally
invariant solutions. For example, solutions invariant under a combination of conformal
generators $C_0, C_1, C_2$, namely $K=\alpha _0 C_0+\alpha _1 C_1+
\alpha _2 C_2$ are provided by the following reduction formula
\begin{equation}\label{ansatz}
u=r^{-1}f(\omega ),\quad \omega =r^{-2}(\beta _0 t+\beta _1 x+\beta _2
y),\quad r^2=t^2-x^2-y^2
\end{equation}
with constants satisfying
$$\beta _2 \alpha _2+\beta _1 \alpha _1-\beta _0 \alpha _0=0.$$
Substitution of \eqref{ansatz} into \eqref{1.2} gives rise to the
second order ODE
\begin{equation}
\beta ^2 f''+a f^5=0,\qquad \beta ^2=\beta _2^2+\beta _1^2-\beta _0^2.
\end{equation}
Again, by a change of dependent variable this equation can be 
transformed to the elliptic function equation.
On the other hand conformal symmetry allows for new solutions to be
produced from Poincar\'e and similitude invariant solutions.
More precisely, whenever $f(\tilde r)$ is a solution
to \eqref{1.2}, so is
$$u({\bf r})=\sigma ^{-1/2} f({\bf \tilde r}),\quad
\sigma =1-{\boldsymbol{\theta}}.{\bf r}+\theta ^2 r^2,\quad
\tilde {\bf r}=\sigma ^{-1}({\bf r}-{\boldsymbol{\theta}}),
\quad {\bf r}=(t,x,y)$$
where $\boldsymbol{\theta}=(\theta _0,\theta _1,\theta _1)$ are group parameters.

\section{Conclusions}
In this paper we combine group theory with singularity analysis to
obtain exact solutions to a conformally invariant nonlinear
Klein-Gordon equation arising in classical and quantum field theory in
(2+1)-dimensions.
We first classify symmetry reductions of the equation
based on a subgroup classification, rather than to
choose intuitively obvious subgroups. In subsection \ref{pdered} using
the subgroups of the similitude group (i.e. the Poincar\'e group extended
by dilations) we list all
the reduced equations in two variables. In subsection \ref{odered} all 
possible symmetry
reductions to first and second order ordinary differential equations are
given.
In the final section solutions of the reduced \ode s are discussed. In
particular, it has been shown that, whenever they pass the Painlev\'e
test, they can be transformed to the equations for  elliptic
functions. Hence, in general, we integrate them in terms of  Jacobi
elliptic functions. As the limiting cases of these functions we obtain
interesting elementary solutions which may be periodic or localized.
In cases
when the reduced second order equations do not have the Painlev\'e property
we reduce
them to Abel type equations of the second kind. A well-known
characteristic of them is
that it is exceptionally that they are integrable with quadratures. That
is why it is no surprise that the ones appearing in the present paper
are far from being integrated by standard methods and they often lack
a symmetry which will enable to find suitable coordinates reducing them to
quadratures.
Note that they may have movable branch points other than movable poles.
This means that an Abel equation does not have  the Painlev\'e property.


\end{document}